\documentclass{article}

\usepackage{newsprocl}
\usepackage[italian,french,german,british]{babel}
\usepackage{amsmath}\usepackage{amssymb}
\usepackage[dvipdf]{graphicx}
\usepackage[numbers,sort&compress]{natbib}
\usepackage{url}
\usepackage{bm}
\newcommand{\stgreek}{%
\DeclareFontFamily{U}{eur}{\skewchar\font'177}%
\DeclareFontShape{U}{eur}{m}{n}{<-6> eurm5 <6-8> eurm7 <8-> eurm10}{}%
\DeclareFontShape{U}{eur}{b}{n}{<-6> eurb5 <6-8> eurb7 <8-> eurb10}{}%
\DeclareSymbolFont{sigreek}{U}{eur}{m}{n}%
\DeclareSymbolFont{boldsigreek}{U}{eur}{m}{n}%
\SetSymbolFont{sigreek}{bold}{U}{eur}{b}{n}
\DeclareSymbolFontAlphabet{\mathgr}{sigreek}%
\DeclareMathSymbol{\rpi}{\mathord}{sigreek}{"19}
\DeclareMathSymbol{\rde}{\mathord}{sigreek}{"40}
\DeclareMathSymbol{\rdelta}{\mathord}{sigreek}{"0E}
\DeclareMathSymbol{\rvarepsilon}{\mathord}{sigreek}{"22}
\DeclareMathSymbol{\rmu}{\mathord}{sigreek}{"16}
\DeclareMathSymbol{\rzeta}{\mathord}{sigreek}{"10}
}

                          \stgreek
\newcommand{\delt}{\rdelta}

\newcommand{\RR}{\mathbb{R}}

\DeclareMathOperator{\tr}{tr}
\newcommand{\tran}{^{\mathsf{T}}}

\newcommand{\cond}{\mathpunct{|}}
\newcommand{\mult}{\times}
\newcommand{\inn}{\cdot}
\newcommand{\defin}{\stackrel{\text{\tiny def}}{=}}
\renewcommand{\le}{\leqslant}
\DeclareMathDelimiter{\lclose}{\mathopen}{operators}{"5B}{largesymbols}{"02}
\DeclareMathDelimiter{\rclose}{\mathclose}{operators}{"5D}{largesymbols}{"03}
\DeclareMathDelimiter{\lopen}{\mathopen}{operators}{"5D}{largesymbols}{"03}
\DeclareMathDelimiter{\ropen}{\mathclose}{operators}{"5B}{largesymbols}{"02}

\newcommand{\set}[1]{\{#1\}}

\newcommand{\pr}{P}

\newcommand{\prop}[1]{\emph{\textsf{`#1'}}}
\newcommand{\degree}{^{\mathord{\circ}}}
\newcommand{\etc}{{etc.}}
\newcommand{\ie}{{i.e.}}
\newcommand{\eg}{{e.g.}}
\newcommand{\viz}{{viz.}}
\newcommand{\cf}{{cf.}}

\newcommand{\etal}{{et al.}}
\newcommand{\qed}{{q.e.d.}}
\DeclareMathOperator{\rank}{rank}
\newcommand{\mprop}[1]{\textsf{\emph{#1}}}
\newcommand{\zM}{M}
\newcommand{\zss}{\mprop{S}}
\newcommand{\zS}{s}
\newcommand{\zs}{\bm{s}}
\newcommand{\zmm}{\mprop{M}}
\newcommand{\zrr}{\mprop{R}}
\newcommand{\zR}{r}
\newcommand{\zr}{\bm{r}}
\newcommand{\zL}{L}
\newcommand{\zP}{\bm{p}}
\newcommand{\zp}{p}
\newcommand{\zK}{K}
\newcommand{\zT}{\bm{t}}
\newcommand{\zt}{t}
\newcommand{\zU}{\bm{u}}
\newcommand{\zu}{u}
\newcommand{\zA}{\bm{a}}
\newcommand{\zB}{\bm{b}}
\newcommand{\zC}{\bm{c}}
\newcommand{\zD}{\bm{d}}
\newcommand{\zV}{\bm{v}}
\newcommand{\zW}{\bm{w}}
\newcommand{\zX}{\bm{x}}
\newcommand{\zY}{\bm{y}}
\newcommand{\zI}{I}
\newcommand{\zd}{D}
\newcommand{\zN}{N}
\newcommand{\zrho}{\Hat{\bm{\rho}}}
\newcommand{\zpov}{\Hat{\bm{\varPi}}}
\newcommand{\zbas}{\Hat{\bm{B}}}
\newcommand{\zpri}{\mprop{Q}}

\begin{document}
\bibliographystyle{apsrevmana}
\title{Probability tables}

\author{Piero~G.~Luca Mana}

\address{Institutionen f{\"o}r mikroelektronik och
  informationsteknik,\\
Kungliga Tekniska H\"ogskolan,\\
\mbox{Isafjordsgatan 22, SE-164\,40 Kista, Sweden}\\
Email: mana@imit.kth.se}

\maketitle
\begin{abstract}
The idea of writing a table of probabilistic data for a quantum or
classical system, and of decomposing this table in a compact way,
leads to a shortcut for Hardy's formalism, and gives new perspectives
on foundational issues.
\end{abstract}



\hspace{\stretch{1}}\begin{minipage}{0.5\columnwidth}
\begin{scriptsize}
\foreignlanguage{german}{--- Aber diess bedeute euch Wille zur
Wahrheit, dass Alles verwandelt werde in
Menschen-\hspace{0pt}Denkbares, Menschen-\hspace{0pt}Sichtbares,
Menschen-\hspace{0pt}F\"uhlbares! Eure
eignen Sinne sollt ihr zu Ende denken!\\
\hspace*{\stretch{1}}Nietzsche}\par
\end{scriptsize}
\end{minipage}
\section{Introduction}\label{sec:intro}
Suppose that we have written, in a sort of table, the statistical
data collected from a group of experiments --- the nature of which
can be classical, quantum, or something else. Suppose that we also
want to store this table's data in a compact way. How could we
proceed?

In this paper it is shown that, given the situation described above,
when we try to store or organise the table's data in a more compact
way we find that real vectors can be associated to preparations and
results, in a way which, for quantum mechanical phenomena, is
essentially the same as Hardy's representation of `states' and
`measurement outcomes'~\cite{Hardy2001}. This curious fact may offer
new points of view for looking at some of the current `foundational'
issues in quantum mechanics.

The ideas here presented are a summary of those developed in
Ref.~\citealp{Mana-ptables}, to which the Reader is referred for
further details. The emphasis in this paper is on the main idea of a
`table decomposition', and on the implications of the latter for
various topics discussed in this Conference.\footnote{Moreover,
many analogies, discussed in Ref.~\citealp{Mana-ptables}, are to be
found between this work and those of
Mielnik~\cite{Mielnik1969,Mielnik1974,Mielnik1976,Mielnik1981},
Foulis and Randall
\etal~\cite{Foulisetal1972a,Randalletal1973b,Foulisetal1978,Randalletal1979,Foulisetal2001},
Barnum~\cite{Barnum2003}, and others.}

\section{Probability tables}\label{sec:decom}
Imagine that we are in a laboratory, performing experiments of
various kinds to study some interesting phenomena; the purpose of the
experiments is to statistically study the correlations among
different kinds of these phenomena. In general, we try to reproduce a
given phenomenon --- either by controllably preparing it at will, or
simply by waiting for its occurrence ---, to observe which
concomitant phenomena, or \emph{results}, occur.

Some experiments present common features: for example, part of the
preparation can be the same for some of them. We separate ideally
each experiment into a \emph{preparation} and an \emph{intervention};
the latter also delimits the kind of results which can be obtained,
which implies that if we are told a result, we know which
intervention was made. 
We then consider a set of preparations and a set of interventions,
with the clause that sensible experiments may be made by combining
each of the preparations with each of the interventions (preparations
or interventions which do not satisfy this condition are set aside
for the moment).\footnote{Note that a preparation does not need to
\emph{temporally precede} an intervention; indeed, these two notions
are meant to have here only a \emph{logical}, \emph{not temporal},
meaning. For example, in quantum-mechanical experiments with
\emph{post-selection}, the preparation is effectively completed
\emph{after} the intervention is made!}

Thus, suppose that we have $\zM$ different preparations $\set{\zss_1,
\dotsc, \zss_{\zM}}$, and a given number of possible interventions
$\set{\zmm_1,\zmm_2,\dotsc,\zmm_k,\dotsc}$, each with a different number
$\zI_{\zmm_k}$ of results $\set{\zrr_1,\zrr_2,\dotsc,\zrr_{\zI_{\zmm_k}}}$
(mutually exclusive and exhaustive%
\footnote{This can always be achieved by grouping in suitable ways
the results, and adding if necessary the result ``\emph{other}''.}),
where the number $\zI_{\zmm_k}$ depends on the particular intervention
$\zmm_k$. The total number of results, counted from all interventions,
is $\zL$.

Through repetitions of the experiments, or through theoretical
assumptions, or just by analogy with other experiments which we have
already seen and which we judge similar to those that are now under
study, we can write down a table $\zP$ with the probabilities that we
assign to every result, for every intervention and preparation. The
table may look like the following:

\begin{center}
\begin{tabular}{cc|ccccccc}
&& $\zss_1$&$\zss_2$& $\zss_3$& $\zss_4$& $\dotso$& $\zss_{\zM}$\\
\hline
$\zmm_1$&
\begin{tabular}{c}$\zrr_1$ \\ $\zrr_2$\end{tabular}& 
\begin{tabular}{c}$\zp_{11}$\\ $\zp_{21}$\end{tabular}&
\begin{tabular}{c}$\zp_{12}$\\ $\zp_{22}$\end{tabular}&
\begin{tabular}{c}$\zp_{13}$\\ $\zp_{23}$\end{tabular}&
\begin{tabular}{c}$\zp_{14}$\\ $\zp_{24}$\end{tabular}&
\begin{tabular}{c}$\dotso$\\ $\dotso$\end{tabular}&
\begin{tabular}{c}$\zp_{1\zM}$\\ $\zp_{2\zM}$\end{tabular}
\\
\hline
$\zmm_2$&
\begin{tabular}{c}$\zrr_3$ \\ $\zrr_4$\\ $\zrr_5$\end{tabular}& 
\begin{tabular}{c}$\zp_{31}$\\ $\zp_{41}$\\ $\zp_{51}$\end{tabular}&
\begin{tabular}{c}$\zp_{32}$\\ $\zp_{42}$\\ $\zp_{52}$\end{tabular}&
\begin{tabular}{c}$\zp_{33}$\\ $\zp_{43}$\\ $\zp_{53}$\end{tabular}&
\begin{tabular}{c}$\zp_{34}$\\ $\zp_{44}$\\ $\zp_{54}$\end{tabular}&
\begin{tabular}{c}$\dotso$\\ $\dotso$\\ $\dotso$\end{tabular}&
\begin{tabular}{c}$\zp_{3\zM}$\\ $\zp_{4\zM}$\\ $\zp_{5\zM}$\end{tabular}
\\
\hline
$\zmm_3$&
\begin{tabular}[t]{c}$\zrr_6$ \\ $\dotso$\\ $\zrr_\zL$\end{tabular}& 
\begin{tabular}[t]{c}$\zp_{61}$\\ $\dotso$\\ $\zp_{\zL1}$\end{tabular}&
\begin{tabular}[t]{c}$\zp_{62}$\\ $\dotso$\\ $\zp_{\zL2}$\end{tabular}&
\begin{tabular}[t]{c}$\zp_{63}$\\ $\dotso$\\ $\zp_{\zL3}$\end{tabular}&
\begin{tabular}[t]{c}$\zp_{64}$\\ $\dotso$\\ $\zp_{\zL4}$\end{tabular}&
\begin{tabular}[t]{c}$\dotso$\\ $\dotso$\\ $\dotso$\end{tabular}&
\begin{tabular}[t]{c}$\zp_{6\zM}$\\ $\dotso$\\ $\zp_{\zL\zM}$\end{tabular}
\end{tabular}
\end{center}

The table, which may be called a `\emph{probability table}', has a
column for each preparation and a group of rows for each
intervention, and these rows are the possible results of the
intervention. The table entry $\zp_{ij}$ is the probability
of obtaining the result $\zrr_i$ for the intervention $\zmm_{k_i}$ and
the preparation $\zss_j$.  For example, the entry $(4,3)$ is the
probability $\zp_{4\,3}$ the we assign to obtaining the result
$\zrr_4$, among the possible results $\set{\zrr_3,\zrr_4,\zrr_5}$,
for the intervention $\zmm_2$ and the preparation $\zss_3$.
Preparations and results can be listed and rearranged in any desired
way in the table. Such a table would very likely have a large number
of rows and columns, \ie, the numbers $\zL$ and $\zM$ are likely to
be very large.


Now, suppose that we seek a more compact way to write down and store
the probability data collected in the table $\zP$. We note that the
table is really just an $\zL \times \zM$ rectangular matrix, and as
such it has a rank $\zK$, \viz, the minimum number of linearly
independent rows or columns:
\begin{equation}\label{eq:rank}
\zK\defin\rank\zP \le \min\set{\zL,\zM}.
\end{equation}
It follows from linear algebra that $\zP$ can be written as the
product of an $\zL \times \zK$ matrix $\zT$ and a $\zK \times \zM$
matrix $\zU$:%
\footnote{This is equivalent to the fact that a linear map
$\zp\colon\RR^{\zM} \to \RR^{\zL}$ of rank $\zK\defin\dim
\zp\bigl(\RR^{\zM}\bigr)$ can be obtained as the composition
$\zp=\zt\circ\zu$ of a surjective map $\zu\colon \RR^{\zM} \to
\zp\bigl(\RR^{\zM}\bigr)$ and an injective map $\zt\colon
\zp\bigl(\RR^{\zM}\bigr) \to \RR^{\zL}$.}
\begin{equation}
\label{eq:decomp1}
\zP =\zT\,\zU,
\end{equation}
or
\begin{equation}
\left(\begin{smallmatrix}
\zp_{11} & \dots &\zp_{1j} &\dots&\zp_{1\zM} \\
\dots&\dots&\dots&\dots&\dots\\
\zp_{i1} & \dots &\zp_{ij} &\dots&\zp_{i\zM} \\
\dots&\dots&\dots&\dots&\dots\\
\zp_{\zL 1} & \dots &\zp_{\zL j} &\dots&\zp_{\zL\zM}
\end{smallmatrix}\right)  
%
= \left(\begin{smallmatrix}
\zr_1\tran \\
\dots\\
\zr_i\tran \\
\dots\\
\zr_{\zL}\tran 
\end{smallmatrix}\right) 
\left(\begin{smallmatrix}
\zs_1 & \dots &\zs_j &\dots&\zs_{\zM} \\
\end{smallmatrix}\right) 
\end{equation}
In the last equation, the matrix $\zT$ has been written as a block of
row vectors $\zr_i\tran$, and the matrix $\zU$ as a block of
column vectors $\zs_i$. In this decomposition, the element $\zp_{ij}$
of $\zP$ is then given by the matrix product of the row vector
$\zr_i\tran$ with the column vector $\zs_j$:
\begin{equation}
\label{eq:baserule}
\zp_{ij}=\zr_i\tran \zs_j=\zr_i \inn \zs_j,
\end{equation}
where, in the last expression, $\zr_i$ and $\zs_j$ are considered as
vectors in $\RR^{\zK}$, so that the matrix product is equivalent to
the scalar product. It will be shown in a moment that the
decomposition is always effective in reducing the number of data of
the table.

The result is that we can associate \emph{vectors} $\set{\zs_j}$ and
$\set{\zr_i}$ in $\RR^{\zK}$, for some $\zK$, to the preparations and
the intervention results for the table, and the relative
probabilities are given by their \emph{scalar product}:
\begin{equation}
\label{eq:repres}
\zp_{ij}
\equiv
\zr_i \inn \zs_j.
\end{equation}
These vectors can be called \emph{preparation vectors} and
\emph{(intervention-)result vectors}, or, in general, \emph{(table)
vectors}.

It should be remarked immediately that we are not postulating any
kind of physical property in Eq.~\eqref{eq:repres}; even less have we
found one. We have only decided to represent and store a collection
of numbers (which do have physical significance) in an alternative
way. Note, in particular, that the numerical values of the vectors
$\set{\zs_j}$ and $\set{\zr_i}$, as well as their dimension $\zK$,
depend on the \emph{whole} collection of probabilities
$\set{p_{ij}}$: if one of these is changed, then $\zK$ and all the
vectors will in general change.

The meaning of the representation~\eqref{eq:repres} shall be
discussed in a moment, but let us study the decomposition in more
detail first, in order to have a clearer idea of how the
table vectors originate.

The matrices $\zT$ and $\zU$ are not uniquely determined from the
decomposition~\eqref{eq:decomp1}, so that there is some freedom in
choosing their form. The fact that $\rank\zP=\zK$, implies that there
exists a square $\zK\times\zK$ submatrix $\zA$, obtained from $\zP$
by suppressing $(\zL-\zK)$ rows and $(\zM-\zK)$ columns, such that
$\det\zA\neq 0$. It is always possible to rearrange the rows and the
columns of the table $\zP$ so that this non-singular submatrix is the
one formed by the first $\zK$ rows and $\zK$ columns. After this
rearrangement,
$\zP$ can be written in the following block form:
\begin{equation}
\label{eq:blockformA}
\zP=\begin{pmatrix}\zA&\zB\\ \zC & \zD \end{pmatrix}\quad
\text{with $\det\zA\neq 0$},
\end{equation}
where $\zB$, $\zC$, and $\zD$ are of order $\zK\times(\zM-\zK)$,
$(\zL-\zK)\times\zK$, and $(\zL-\zK)\times(\zM-\zK)$ respectively.

By writing also the matrices $\zT$ and $\zU$ in block form
\begin{equation}
\label{eq:blockformTU}
\zT=\begin{pmatrix}\zV\\ \zW \end{pmatrix},\quad
\zU=\begin{pmatrix}\zX & \zY \end{pmatrix},
\end{equation}
where the orders of $\zV$, $\zW$, $\zX$, and $\zY$ are
$\zK\times\zK$, $(\zL-\zK)\times\zK$, $\zK\times\zK$, and
$\zK\times(\zM-\zK)$ respectively, we can rewrite
the decomposition equation~\eqref{eq:decomp1} as
\begin{gather}
\begin{pmatrix}\zA&\zB\\ \zC & \zD \end{pmatrix} =
\begin{pmatrix}\zV\\ \zW \end{pmatrix} 
\begin{pmatrix}\zX & \zY \end{pmatrix},\label{eq:blockdecompmatr}\\
\intertext{or}
\begin{aligned}
\zA&=\zV \zX, &
\zB&=\zV \zY,&
\zC&=\zW \zX, &
\zD&=\zW \zY,
\end{aligned}\label{eq:blockdecomp}
\end{gather}
which has the solution\footnote{Note that $\zD=\zC \zA^{-1} \zB$,
\ie, the submatrix $\zD$ of $\zP$ is completely determined by the
other submatrices $\zA$, $\zB$, and $\zC$, because $\rank\zP=\zK$.}
\begin{equation}
\begin{aligned}
\det\zX&\neq 0,&
\zY&=\zX \zA^{-1}\zB, &\zV&=\zA \zX^{-1},& 
\zW&=\zC \zX^{-1}, 
\end{aligned}\label{eq:solutiondecomp}
\end{equation}
which in terms of the result and preparation vectors is
\begin{subequations}\label{eq:solutiondecompmatr}
\begin{gather}
\begin{aligned}
(\zs_1\;\dots\;\zs_\zK) &=\zX, &\label{eq:matrixX}
(\zs_{\zK+1}\;\dots\;\zs_\zM) &=\zX \zA^{-1} \zB,
\end{aligned}
\\
\begin{aligned}
\begin{pmatrix}\zr_1\\ \dots\\ \zr_\zK\end{pmatrix}
&=\zA \zX^{-1}, &
\begin{pmatrix}\zr_{\zK+1}\\ \dots\\ \zr_\zL\end{pmatrix}
&=\zC \zX^{-1}.
\end{aligned}
\end{gather}
\end{subequations}
The square matrix $\zX = (\zs_1\;\dots\;\zs_{\zK})$ is undetermined
except for the condition of being non-singular; this corresponds to
the freedom of choosing $\zK$ \emph{basis vectors} in $\RR^{\zK}$ as
those associated to the first $\zK$ preparations\footnote{There is
the alternative option of choosing the vectors associated to the
first $\zK$ results; this corresponds to solving
Eq.~\eqref{eq:blockdecomp} in terms of $\zV$.}
$\zss_1,\dotsc,\zss_{\zK}$, which can then be called \emph{basis
preparation-vectors}.  Note that for a given probability table $\zP$
the set of basis preparation-vectors is in general non-unique,
because $\zP$ possesses in general many submatrices like $\zA$ with
rank $\zK$.

We can now check whether the above decomposition effectively reduces
the number of data to be stored. The table $\zP$ has $\zL\times\zM$
entries. The matrices $\zX$, $\zY$, $\zV$, $\zW$ (or equivalently the
collection of vectors $\set{\zs_j}$ and $\set{\zr_i}$) have in total
$\zK\times(\zL+\zM)$ entries; however, we are free to choose a
`canonical' form for the non-singular $\zK\times\zK$ matrix $\zX$
(\eg, the identity matrix) once and for all, for all probability
tables (this corresponds to a standard choice of basis vectors in
$\RR^\zK$). This implies that we are indeed left with only
$\zK\times(\zL+\zM) -\zK^2$ numbers. It is then simple to verify
that, from $\zK\le \min\set{\zL,\zM}$, we have
\begin{equation}
\label{eq:lessdata}
\zK\times(\zL+\zM) -\zK^2\le\zL\times\zM.
\end{equation}
Thus the decomposition of the probability table into two sets of
vectors is indeed a more compact way to write down and store the
probability data.

\subsection{A numerical and graphical example}
\label{sec:exampletable}
Imagine that, with the various apparatus in our laboratory, we can
make seven different preparations $\set{\zss_1,\dotsc,\zss_7}$ and
perform three different interventions $\set{\zmm_1,\zmm_2,\zmm_3}$, each
having two results. The probabilities that we assign to the various
results are given in the following table:

\begin{center}
\begin{tabular}{cc|ccccccc}
&& $\zss_1$&$\zss_2$& $\zss_3$&$\zss_4$&$\zss_5$&$\zss_6$&$\zss_7$\\
\hline
$\zmm_1$
&\begin{tabular}{c}$\zrr_1$\\$\zrr_2$\end{tabular}
&\begin{tabular}{c}1\\0\end{tabular}
&\begin{tabular}{c}$\tfrac{1}{2}$\\$\tfrac{1}{2}$\end{tabular}
&\begin{tabular}{c}0\\1\end{tabular}
&\begin{tabular}{c}$\tfrac{1}{2}$\\$\tfrac{1}{2}$\end{tabular}
&\begin{tabular}{c}$\tfrac{3}{4}$\\$\tfrac{1}{4}$\end{tabular}
&\begin{tabular}{c}$\tfrac{1}{2}$\\$\tfrac{1}{2}$\end{tabular}
&\begin{tabular}{c}$\tfrac{3}{4}$\\$\tfrac{1}{4}$\end{tabular}
\\[1ex]
\hline
$\zmm_2$
&\begin{tabular}{c}$\zrr_3$\\$\zrr_4$\end{tabular}
&\begin{tabular}{c}$\tfrac{1}{2}$\\$\tfrac{1}{2}$\end{tabular}
&\begin{tabular}{c}1\\0\end{tabular}
&\begin{tabular}{c}$\tfrac{1}{2}$\\$\tfrac{1}{2}$\end{tabular}
&\begin{tabular}{c}0\\1\end{tabular}
&\begin{tabular}{c}$\tfrac{3}{4}$\\$\tfrac{1}{4}$\end{tabular}
&\begin{tabular}{c}$\tfrac{1}{2}$\\$\tfrac{1}{2}$\end{tabular}
&\begin{tabular}{c}$\tfrac{1}{2}$\\$\tfrac{1}{2}$\end{tabular}
\\
\hline
$\zmm_3$
&\begin{tabular}{c}$\zrr_5$\\$\zrr_6$\end{tabular}
&\begin{tabular}{c}1\\0\end{tabular}
&\begin{tabular}{c}1\\0\end{tabular}
&\begin{tabular}{c}1\\0\end{tabular}
&\begin{tabular}{c}1\\0\end{tabular}
&\begin{tabular}{c}1\\0\end{tabular}
&\begin{tabular}{c}1\\0\end{tabular}
&\begin{tabular}{c}1\\0\end{tabular}
\end{tabular}
\end{center}

\begin{figure}[!h]
\begin{center}
\includegraphics[width=.7\columnwidth]{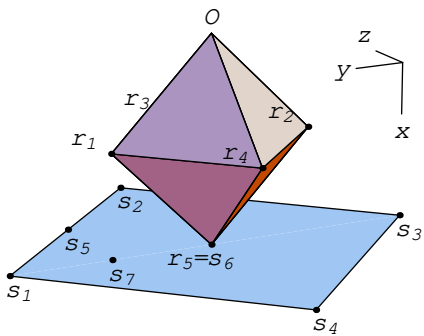}
\caption{Preparation and result vectors (and their convex hulls) in
$\RR^3$, from the example. $O$ is the origin; $x$, $y$, $z$ are the
directions of the Cartesian axes.} \label{fig:example}
\end{center}
\end{figure}

The matrix corresponding to this probability table
has rank $\zK=3$, and the first $3\times3$ submatrix $\zA$ is indeed
already non-singular, without the need of rearranging the rows or
columns of the original table. Proceeding as in the decomposition
equations~\eqref{eq:blockdecomp} and~\eqref{eq:solutiondecomp} with
the following choice for the matrix $\zX$:
\begin{equation}
\label{eq:numermatrX}
\zX\equiv\begin{pmatrix}
1&1&1 \\
1&0&-1 \\
0&1&0
\end{pmatrix},
\end{equation}
we finally obtain the following preparation and result vectors:
\begin{align}
&\begin{aligned}
\zs_1&=(1,1,0),\\ 
\zs_2&=(1,0,1), \\ 
\zs_3&=(1,-1,0), \\
\zs_4&=(1,0,-1), \\
\zs_5&=(1,\tfrac{1}{2},\tfrac{1}{2}), \\ 
\zs_6&=(1,0,0), \\
\zs_7&=(1,\tfrac{1}{2},0), 
\end{aligned}&&
\begin{aligned}
\zr_1&=(\tfrac{1}{2},\tfrac{1}{2},0),\\
\zr_2&=(\tfrac{1}{2},-\tfrac{1}{2},0 ),\\
\zr_3&=(\tfrac{1}{2},0,\tfrac{1}{2}),\\
\zr_4&=(\tfrac{1}{2},0,-\tfrac{1}{2}),\\
\zr_5&=(1,0,0 ),\\
\zr_6&=\bm{O}=(0,0,0 ),\\
\end{aligned}
\end{align}
represented as points of $\RR^3$ in Fig.~\ref{fig:example}.

We notice that all preparation vectors lie on the same plane ($x=1$);
this is a general property of any probability table, which derives
from the fact that the results of an intervention are mutually
exclusive and exhaustive; this also implies that an intervention's
result-vectors always sum up to the same vector~\cite{Mana-ptables}: in
this case, $\zr_1+\zr_2 = \zr_3+\zr_4 = \zr_5 +\zr_6 = (1,0,0)$.

\section{Relation to quantum mechanics}
\label{sec:propandQM}
Probability tables like the one illustrated above can be made,
\emph{in particular}, for phenomena concerning classical or quantum
systems. Indeed, the resulting association of vectors to preparations
and results is analogous to that introduced by
Hardy~\cite{Hardy2001}, through a different line of reasoning, for
`states' and `measurement outcomes' of classical and quantum systems.

For example, imagine how a table for a two-level quantum system would
appear; consider for concreteness the polarisation of a single
photon. We can prepare the photon with different polarisation
directions and with different degrees of polarisation; hence the
table has in the limit a continuum of columns, one for each
preparation. Analogously, we can perform some interventions on the
photon by placing various polarisation filters on its way,
controlling if it is absorbed or not; in the limit, the table has
also a continuum of rows, one for each intervention result. Some
entries of the table would look like the following (the meaning of
the symbols should be self-evident):
\begin{center}
\begin{tabular}{cc|ccccc}
&& $\dots$ &$\zss_{0\degree}$& $\dots$&$\zss_{45\degree}$& $\dots$\\
\hline
$\dots$&$\dots$&$\dots$&$\dots$&$\dots$&$\dots$&$\dots$\\
\hline
$\zmm_{45\degree}$
&\begin{tabular}{c}$\zrr_{\text{out}}^{45\degree}$\\
$\zrr_{\text{abs}}^{45\degree}$\end{tabular}
&\begin{tabular}{c}$\dotso$\\$\dotso$\end{tabular}
&\begin{tabular}{c}$\tfrac{1}{2}$\\$\tfrac{1}{2}$\end{tabular}
&\begin{tabular}{c}$\dotso$\\$\dotso$\end{tabular}
&\begin{tabular}{c}1\\0\end{tabular}
&\begin{tabular}{c}$\dotso$\\$\dotso$\end{tabular}
\\
\hline
$\dots$&$\dots$&$\dots$&$\dots$&$\dots$&$\dots$&$\dots$\\
\hline
$\zmm_{60\degree}$
&\begin{tabular}{c}$\zrr_{\text{out}}^{60\degree}$\\
$\zrr_{\text{abs}}^{60\degree}$\end{tabular}
&\begin{tabular}{c}$\dotso$\\$\dotso$\end{tabular}
&\begin{tabular}{c}$\tfrac{1}{4}$\\$\tfrac{3}{4}$\end{tabular}
&\begin{tabular}{c}$\dotso$\\$\dotso$\end{tabular}
&\begin{tabular}{c}$0.933$\\$0.067$\end{tabular}
&\begin{tabular}{c}$\dotso$\\$\dotso$\end{tabular}
\\
\hline
$\dots$&$\dots$&$\dots$&$\dots$&$\dots$&$\dots$&$\dots$
\end{tabular}
\end{center}


Such a table has, notwithstanding the limiting infinite size, rank
$\zK=4$, and so we can associate to every preparation and to every
result a $4$-dimensional vector. The preparation vectors, however,
lie on a $3$-dimensional (affine) hyper-plane, as in the numerical
example previously discussed. It can be shown
indeed~\cite{Hardy2001,Mana-ptables} that the resulting set of
preparation vectors is equivalent to the standard Bloch-sphere for
two-level quantum systems.

This was just an example, but all quantum-mechanical concepts like
density matrix, positive-operator-valued measure, and completely
positive map can be expressed in an equivalent table-vector
formalism~\cite{Hardy2001,Mana-ptables} (\eg, the relation with the
trace rule is quickly shown in the appendix).


\vspace{-\medskipamount}
\section{Discussion}
\label{sec:discussion1}

The following discussion concentrates on the possible relations among
the ideas hitherto presented and ideas presented by other authors in
this Conference.  The Reader is referred to
Ref.~\citealp{Mana-ptables} for a more general and
detailed discussion.

\vspace{-\medskipamount}\subsection{On the probabilities in the table and `quantum logics'}
\label{sec:probab}
In Sect.~\ref{sec:decom}, we quickly introduced the preparations,
interventions, and results $\set{\zss_j}$, $\set{\zmm_k}$, $\set{\zrr_i}$, and
the probabilities $\set{\zp_{ij}}$ which make up the table. Let us define
them more clearly.

Symbols like $\zss_i$ and $\zmm_k$ represent propositions which
together describe an actual, well-defined procedure to set up an
experiment, \eg\ $\zss_i=$\prop{The diode laser is put on the table
is such and such position, with a beam splitter in front of it in
such and such position\ldots}\ \etc,\footnote{Another example:
\prop{Wait until such and such happens, then\ldots}\ \etc}\ and
$\zmm_k=$\prop{Such and such vertical filter is placed in such and
such place, and the detector is placed behind it\ldots}\ \etc{} The
separation of the experiment's description into the two propositions
is not unique, and indeed more kinds of separations can be
considered.\footnote{In Ref.~\citealp{Mana-ptables}, this fact is used
as the starting point to define the concept of
\emph{transformation}.}

A symbol like $\zrr_j$ represents a proposition which describes the
results of an experiment, \eg\ $\zrr_j=$\prop{The detector does not
click}; it is then clear that it depends on the particular experiment
being performed.

The probability $\zp_{ij}$ is consequently defined as
\begin{equation}
\label{eq:probasentries}
\zp_{ij}\defin
\pr(\zrr_i\cond \zmm_{k_i}\land
\zss_j\land \zpri),
\end{equation}
where $\zpri$ is a proposition representing the rest of the
experimental details and our prior knowledge.\footnote{A `subjective
Bayesian' can see $\zpri$ as representing the `agent'.} We are thus
considering probability theory as extended
logic~\cite{jeffreys1998,Cox1946,cox1961,jaynes2003}, an approach
which will prove to be, in the following, powerful, flexible, and
intuitive at once.

Usually, many repetitions of the same experiment are made and the
relative frequencies of the different results of the intervention are
observed. \emph{In this case, given a judgement of total
exchangeability of the experiment's repetitions, the probability is
practically equal to the observed frequency of the result}, thanks to
de~Finetti's representation theorem~\cite{Definetti1964}. 
But the probability can also be assigned on grounds of similarity
with other experiments, or just by theoretical assumptions.

Note that the table is silent with regard to the probabilities for
the preparations or the interventions, $\pr(\zss_j\cond
\zpri)$ and $\pr(\zmm_{k}\cond \zpri)$. In a given
`situation' $\zpri$, if \emph{we} decide which preparation and
intervention to perform, say $\zss_5$ and $\zmm_7$, then these
probabilities are $\pr(\zss_i\cond\zpri)=\delt_{i,5}$ and
$\pr(\zmm_k\cond\zpri)=\delt_{k,7}$ of course (which amounts
to saying that ``we know what we're doing''). If it is someone else
who is deciding the particular preparation and intervention, then the
probabilities must be assigned in some other way, \eg\ by asking
``which preparation are you making?'', or by using some other
knowledge, and they will in general differ from $0$ and $1$. The
difference between these probabilities and those of 
Eq.~\eqref{eq:probasentries} is somehow analogous to the difference between
initial conditions and equations of motion in classical mechanics:
the theory concerns only the latter, while the former has to be
specified on a case-by-case basis.

The probabilities of the results of a given intervention and a given
preparation form a probability distribution, because the results
were arranged so as to be mutually exclusive and exhaustive. This
implies that, for two results $\zrr_{k}$ and $\zrr_{k}'$ of a given
intervention $\zmm_k$ and for a given preparation $\zss_j$, we also have
the trivial identity
\begin{equation}
\label{eq:trivialprobsum}
\begin{split}
\pr(\zrr_{k}\lor\zrr_{k}'\cond
\zmm_{k}\land \zss_j\land \zpri)&=
\pr(\zrr_{k}\cond \zmm_{k}\land \zss_j\land
\zpri)+
 \pr(\zrr_{k}'\cond \zmm_{k}\land
\zss_j\land \zpri),
\\
&=(\zr_{k}+\zr_{k}')\inn \zs_j.\raisetag{3ex}
\end{split}
\end{equation}

But with probability theory as logic we can also evaluate, for a
given preparation $\zss_j$, the disjoint probability for the results
$\zrr'$ and $\zrr''$ of two \emph{different} interventions $\zmm'$ and
$\zmm''$, just using the product and sum rules:
\begin{equation}
\label{eq:nontrivsum}
\begin{split}
\pr[\zrr'\lor\zrr''\cond (\zmm'\lor\zmm'') \land \zss_j\land \zpri]&=
\pr[(\zrr'\land\zmm')\lor (\zrr''\land \zmm'')
\cond \zss_j\land \zpri],\\
&=
\begin{aligned}[t]
&\pr(\zrr'\cond \zmm'\land \zss_j\land \zpri)\times
\pr(\zmm'\cond \zss_j\land \zpri)+\mbox{}\\
&\pr(\zrr''\cond \zmm''\land \zss_j\land \zpri)\times \pr(\zmm''\cond
\zss_j\land \zpri),
\end{aligned}\\
&=[\zr' \pr(\zmm'\cond \zss_j\land \zpri)+ \zr'' \pr(\zmm''\cond
\zss_j\land \zpri)] \inn \zs_j,
\end{split}
\end{equation}
where it is assumed that $\pr(\zmm'\lor\zmm''\cond
\zss_j\land \zpri)=1$, \ie, we are sure that one or
the other intervention was performed.

The content of the formula above is intuitive: the occurrence of the
result $\zrr'$ implies that the intervention $\zmm'$ was performed,
and so analogously for the result $\zrr''$ and the intervention
$\zmm''$. Then the probability of getting the one or the other result
depends in turn on the probability that the one or the other
intervention was made, hence these probabilities appear in the last line
of the above equation. However, as already said, the probabilities of
these interventions are \emph{not} contained in the table, but must
be given on a case-by-case basis.

As a result, the two disjoint probabilities in
Eqs.~\eqref{eq:trivialprobsum} and~\eqref{eq:nontrivsum} `behave'
differently, and the reason for this is intuitively clear. However,
we can partially trace in this fact the source of much research and
discussion on partially ordered lattices and quantum logics for the
set of intervention results~\cite{jauch1973,piron1976,wilce-ptables}.
Roughly speaking, the point is that, for a classical system, there is
the theoretical possibility of joining all possible interventions
(measurements) in a single ``total intervention'': the table
associated to a classical system can then be considered as having
only one intervention; thus one needs never consider the case of
Eq.~\eqref{eq:nontrivsum}. However, such ``total intervention'' is
excluded in quantum mechanics, and one is thus forced to consider the
case of Eq.~\eqref{eq:nontrivsum}.

From the point of view of probability theory as logic, instead, there
is no need for non-Boolean structures thanks to the possibility of
changing and adapting, by means of Bayes' theorem, the \emph{context}
(also called `prior knowledge'~\cite{jaynes2003} or
`data'~\cite{jeffreys1998}) of a probability, \ie, the proposition to
the right of the conditional symbol `$\cond$'.\footnote{In contrast,
`Kolmogorovian' probability, with its scarce flexibility with respect
to contexts, deals with these features with difficulty. Loubenets'
efforts~\cite{Loubenets2003,Loubenets2003b} are directed to ameliorating this
situation.}

\subsection{Unknown preparations and their `tomography'}
\label{sec:tomography}
On the other hand, we may have the following scenario: we have
repeated instances of a given preparation, but we do not know which.
By performing interventions on these instances and observing the results,
we can estimate which preparation is being made. Introducing the
proposition $\zd$ representing the results thus obtained, we have
\begin{equation}
\label{eq:invinfer}
\pr(\zss_j\cond \zd\land\zpri)=
\frac{\pr(\zd\cond \zss_j\land\zpri)\,
\pr(\zss_j\cond \zpri)}{%
\sum_j \pr(\zd\cond \zss_j\land\zpri)\,
\pr(\zss_j\cond \zpri)}.
\end{equation}
This is a standard ``inverse-inference'' result of Bayesian
analysis~\cite[Ch.~4]{jaynes2003}. The probability $\pr(\zd\cond
\zss_j\land\zpri)$ can be written in terms of scalar products
of result and preparation vectors, but the probability distribution
$\set{\pr(\zss_j\cond \zpri)}$ depends on the prior knowledge
that one has in each specific case. If we now want to \emph{predict}
which result will occur in a new intervention $\zmm_k$, we have the
following probability:
\begin{equation}
\label{eq:prediction}
\begin{split}
\pr(\zrr_{i}\cond \zmm_{k_i}\land \zd\land \zpri)
&=\sum_j \pr(\zrr_{i}\cond \zmm_{k_i}\land
\zss_j\land \zpri)\mult \pr(\zss_j \cond \zd\land
\zpri),\\
&=\zr_i\inn\sum_j \zs_j\: \pr(\zss_j \cond \zd\land
\zpri),
\end{split}
\end{equation}
\ie, we can effectively associate the vector
\begin{equation}
\zs_\text{new}\equiv\sum_j \zs_j\,
\pr(\zss_j \cond \zd\land \zpri)\label{eq:newvect}
\end{equation}
to the unknown preparation.

A further kind of scenario is this: we have a brand-new kind of
preparation, \ie, a new phenomenon, still untested. It has \emph{no
place} in our probability table; yet \emph{we think that we could
reserve a new column to it without making substantial changes to our
table's decomposition (by which we mean that the table's rank $\zK$
would not change)}.\footnote{This condition seems to be somehow
related to Fuchs' ``accepting quantum
mechanics''~\cite[Sec.~6]{Fuchs2002}.} Given this, in order to
associate a vector to this new preparation we proceed as in the
preceding scenario, performing interventions and observing results.
The result is that Eqs.~\eqref{eq:invinfer} and~\eqref{eq:newvect}
also apply in this case.

Note that there is no conflict between our talking about ``unknown
preparations'' and Fuchs' criticism of the term `unknown quantum
state'~\cite{Fuchs2002,Vanenketal2002,Vanenketal2002b}. A
preparation, as we have seen, is a well-defined procedure (that can
be shown or described to others, \etc)\ to set up a given physical
situation; on the other hand, Fuchs' meaning of `quantum state' is
`density matrix'~\cite{Fuchs2002,Fuchs-ptables}, which corresponds (more
or less) to the preparation \emph{vector} instead.\footnote{Sadly,
the Janus-faced term `state' is sometimes meant as `density matrix'
and sometimes as a sort of `physical state of affairs', though the
two notions are, of course, quite distinct.} Thus, consider the two
sentences ``It is unknown to me which kind of laser and of beam
splitter are used in this experiment'' and ``I do not know with which
probability the detector behind the vertical filter will click'': the
latter sentence is nonsensical from a Bayesian point of view, because
there are no `unknown' degrees of belief; but the former sentence is
unquestionably meaningful. Thus, even if the \emph{preparation} is
unknown to us, we can \emph{always} associate a
preparation \emph{vector} to it.
\footnote{It would be interesting, indeed, to see more clearly the
connexion between the ``preparation tomography'' illustrated above,
and the `quantum Bayes rule' and `quantum de~Finetti theorem' of
Caves, Fuchs, and
Schack~\cite{Schacketal2001,Cavesetal2002b,Cavesetal2002,Fuchs2002,Vanenketal2002,Vanenketal2002b,schack-ptables};
it seems reasonable to expect that these `quantum' analogues of
Bayesian formulae should have a counterpart for tables concerning
general systems, not only quantum-mechanical ones; the failure of the
`quantum de~Finetti theorem' for quantum mechanics on real Hilbert
spaces is quite interesting from this point of view.}

\subsection{On the content of the table vectors and `quantum states
of knowledge'}
\label{sec:tablevectors}
Some remarks have already been made in Sect.~\ref{sec:decom} after the
derivation of the formula
\begin{equation}
\label{eq:represbis}
\pr(\zrr_i\cond \zmm_{k_i}\land
\zss_j\land \zpri)\equiv
\zr_i \inn \zs_j,\tag{\ref{eq:repres}$_\text{bis}$}
\end{equation}
with regard to the fact that the vectors $\set{\zs_j}$ and
$\set{\zr_i}$ do not have any physical meaning \emph{separately}: a
given preparation vector $\zs_j$ tells us nothing if we do not have
an intervention-result vector ${\zr_i}$; even less if we know nothing
about the set of intervention results. It follows that the
preparation vectors also lack any \emph{probabilistic} meaning
\emph{per se}: they are \emph{not} probabilities \emph{nor}
collections of probabilities --- they are just mathematical objects
which yield probabilities when combined in a given way with objects
of similar kind. This, in particular, is true for density matrices,
as we have seen that they are just a particular case of
preparation vectors. It is slightly incorrect, as well, to say that
probabilities are `encoded' or `contained' in the (quantum)
preparation vectors: rather, they are \emph{parts} of an encoding.

From these considerations, the quantum state (the density matrix)
appears to be \emph{part} of a state of belief, and not the `whole'
state of belief~\cite{Cavesetal2002b,Fuchs2002}. Perhaps the point is
that Caves, Fuchs, and Schack's notion of a `quantum' state of belief
implicitly assumes the \emph{existence} and the \emph{particular
structure} of the whole set of quantum positive-operator-valued
measures (\ie, the interventions).\footnote{However, Caves, Fuchs,
and Schack have the right to be not so pedantic about this point,
because they assume at the start the absolute validity of quantum
mechanics and of its mathematical structure (an assumption not made
in the present work).} This is an important difference from a `usual'
degree of belief $\pr(\mprop{A}\cond \mprop{X})$ which does not need
to be combined with other mathematical objects to reveal its content.


\subsection{Possible applications of the table formalism}
\label{sec:applications}

It has already been remarked that the kind of vector representation
arising from the table decomposition is essentially the same as
Hardy's~\cite{Hardy2001}. The derivation presented here can be seen
as a sort of shortcut for his derivation, but it implies something
more. Hardy supposes that it is possible to represent a preparation
by means of a $\zK$-dimensional vector
with $\zK\le\zL$ because most physical theories have some structure
which relates different measured quantities; but the reasoning behind
the decomposition of Sect.~\ref{sec:decom} shows that this possibility
exists even without a theory that describes the data (indeed, the
question arises: has this possibility any physical meaning at all?)

In any case, the idea of a `probability table' and its decomposition
has probably very little usefulness in experimental applications, but
provides a very simple approach to study the mathematical and
geometrical structures of classical and quantum theories, and offers
a different way to look at their ``foundational'' and
``interpretative'' issues.

This approach is even more general than other standard ones based,
\eg, on $C^*$-algebras, or even
Jordan-Banach-algebras~\cite{Halvorson2003}, or on convex
state-spaces.\footnote{The first two formalisms are included as
particular limiting cases of tables with an uncountably infinite
number of columns and rows. In contrast to the convex-state-space
approach, the probability-table idea does not require that the set of
preparations be necessarily convex (however, the convexity usually
appears in a natural way: see \eg\ Hardy's discussion in
Ref.~\citealp[Sect.~6.5]{Hardy2001}). Moreover, in contrast to all
three mentioned frameworks, it does not assume that the set of
preparations (states) be necessarily the \emph{whole} set of
normalised positive linear functionals of the set of results (POVM
elements), or vice versa for the convex-state-space framework. For
example, (the convex hull of) the set of preparations for the table
of Sect.~\ref{sec:exampletable} is only part of the set of normalised
positive linear functionals of (the convex hull of) the set of
results (the latter would be a square circumscribed on the given
one).} Thus, with the idea of a `probability table' we can very
easily implement `toy theories' or models like those of
Spekkens~\cite{Spekkens2004} and Kirkpatrick~\cite{Kirkpatrick2001}
(\cf\ also the issue raised by Terno~\citep{Terno2004}), which can
then be compared to classical or quantum mechanics using a unique,
common formalism.



\section*{Acknowledgements}
The author would like to thank Gunnar~Bj\"ork, Ingemar Bengtsson,
Christopher Fuchs, Lucien Hardy, \AA{}sa Ericsson, Anders
M\aa{}nsson, and Anna for advice, encouragement, and many useful
discussions.

\appendix

\section{The trace rule}
\label{sec:tracerule}
It is shown that the `scalar product formula', Eq.~\eqref{eq:repres},
includes also the `trace rule' of quantum mechanics (see also
Ref.~\citealp[Sect.~5]{Hardy2001}).

A preparation is usually associated in quantum mechanics to a
\emph{density matrix} $\zrho_j$, and a intervention result to a
\emph{positive-operator-valued-measure element} $\zpov_i$; both are Hermitian
operators in a Hilbert space of dimension $\zN$. The probability of
obtaining the result $\zpov_i$ for a given intervention on preparation
$\zrho_j$ is given by the trace formula
\begin{equation}
\zp_{ij}=\tr\zpov_i \zrho_j.\label{eq:traceform}
\end{equation}
The Hermitian operators form a linear space of \emph{real} dimension
$\zK=\zN^2$; one can choose $\zK$ linearly independent Hermitian
operators $\set{\zbas_k}$ as a basis for this linear space. These can
also be chosen (basically by Gram-Schmidt orthonormalisation) to
satisfy
\begin{equation}
\label{eq:basiskron}
\tr{\zbas_k \zbas_l}=\delt_{kl}.
\end{equation}
Both $\zrho_j$ and $\zpov_i$ can be written as a linear combination of
the basis operators:
\begin{align}
\label{eq:writewithbasis}
\zrho_j&=\sum_{l=1}^{\zK} {\zS_j}^l \zbas_l,&
\zpov_i&=\sum_{k=1}^{\zK} {\zR_i}^k \zbas_k,
\end{align}
where the coefficients ${\zS_j}^l$ and ${\zR_i}^k$ are real. Using
Eqs.~\eqref{eq:writewithbasis} and~\eqref{eq:basiskron} the trace
formula becomes
\begin{equation}
\zp_{ij}=\tr\zpov_i \zrho_j=\sum_{k,l=1}^{\zK} {\zR_i}^k {\zS_j}^l
\tr\zbas_k \zbas_l=
 \sum_{l=1}^{\zK} {\zR_i}^l {\zS_j}^l 
= \zr_i \inn\zs_j,
\label{eq:scalarprform}
\end{equation}
where $\zr_i\defin ({\zR_i}^1,\dotsc,{\zR_i}^{\zK})$ and
$\zs_j\defin ({\zS_j}^1,\dotsc,{\zS_j}^{\zK})$ are vectors in
$\RR^{\zK}$, \qed

\end{document}